\documentclass[aip,apl, amsmath,amssymb, reprint]{revtex4-1}
\usepackage{graphicx}
\usepackage{dcolumn}
\usepackage{bm}
\usepackage{units,libertine,upgreek}
\usepackage[caption=false]{subfig}
\usepackage{siunitx}
\usepackage{color}
\begin{document}
\preprint{AIP/123-QED}
\title{Integrated transition edge sensors on lithium niobate waveguides}
\author{Jan Philipp H{\"o}pker}
\email{jan.philipp.hoepker@upb.de}
\affiliation{Department of Physics, Paderborn University, Warburger Str. 100, 33098 Paderborn, Germany}
\author{Thomas Gerrits}
\affiliation{National Institute of Standards and Technology, 325 Broadway, Boulder, Colorado 80305, USA}
\author{Adriana Lita}
\affiliation{National Institute of Standards and Technology, 325 Broadway, Boulder, Colorado 80305, USA}
\author{Stephan Krapick}
\affiliation{Department of Physics, Paderborn University, Warburger Str. 100, 33098 Paderborn, Germany}
\author{Harald Herrmann}
\affiliation{Department of Physics, Paderborn University, Warburger Str. 100, 33098 Paderborn, Germany}
\author{Raimund Ricken}
\affiliation{Department of Physics, Paderborn University, Warburger Str. 100, 33098 Paderborn, Germany}
\author{Viktor Quiring}
\affiliation{Department of Physics, Paderborn University, Warburger Str. 100, 33098 Paderborn, Germany}
\author{Richard Mirin}
\affiliation{National Institute of Standards and Technology, 325 Broadway, Boulder, Colorado 80305, USA}
\author{Sae Woo Nam}
\affiliation{National Institute of Standards and Technology, 325 Broadway, Boulder, Colorado 80305, USA}
\author{Christine Silberhorn}
\affiliation{Department of Physics, Paderborn University, Warburger Str. 100, 33098 Paderborn, Germany}
\author{Tim J. Bartley}
\affiliation{Department of Physics, Paderborn University, Warburger Str. 100, 33098 Paderborn, Germany}
\date{17 December 2018, Preprint, Submitted to: APL Photonics}
\begin{abstract}
We show the proof-of-principle detection of light at 1550\,nm coupled evanescently from a titanium in-diffused lithium niobate waveguide to a superconducting transition edge sensor. The coupling efficiency strongly depends on the polarization, the overlap between the evanescent field, and the detector structure. We experimentally demonstrate polarization sensitivity of this coupling as well as photon-number resolution of the integrated detector. The combination of transition edge sensors and lithium niobate waveguides can open the field for a variety of new quantum optics experiments.
\end{abstract}
\keywords{Integrated optics, integrated superconducting detectors}
\maketitle

\section{Introduction} \label{kap:introduction}
Integrated photonic circuits are widely used to realize compact and complex quantum optics experiments. They enable scalable creation and processing of quantum states which can be used in communication, computation and simulation protocols to potentially outperform classical systems~\cite{Walmsley}. Lithium niobate is an established platform in the field of classical integrated optics because of its  high second-order susceptibility and electro-optic properties~\cite{Sohler}. Titanium in-diffused waveguides in lithium niobate are surface-guiding and offer low-loss waveguiding in both polarization directions (TE and TM), across a broad frequency range, making them ideal for a range of quantum optical circuits. For example, integrated single-photon sources using parametric down conversion~\cite{Montaut, Krapick} can be combined with electro-optic modulators~\cite{Moeini, Huang} and active or passive routing to enable a variety of applications in quantum optics~\cite{Sharapova}. In addition, low-loss fiber pigtailing, by attaching single-mode fibers directly to the waveguide end-faces (``butt-coupling''), enables high overall system efficiency~\cite{Montaut} and compatibility with the existing communication network infrastructure. 

Until now, detection from lithium niobate chips has been restricted to fiber-coupled detectors, which adds an extra interface and associated losses. Of the various types of  fiber-coupled detectors, those based on the breakdown of superconductivity offer the highest efficiency at telecom wavelengths and can be tailored for low timing-jitter, different photon numbers, or even photon-number resolution~\cite{Migdall, Hadfield, Natarajan, Marsili, Lita, Fukuda, Gerrits2}. However, the positioning of fiber-coupled detectors with respect to an integrated optical circuit is limited, and their scaling towards complex circuitry is challenging. 

Integrated detectors, using the coupling from an evanescent field of a waveguide into an on-chip detector, enable more complex circuitry, as they can be deposited at different positions inside the optical circuit. Furthermore, in an in-line geometry, non-detected (and non-scattered) photons remain inside the waveguide; this geometry potentially allows for further processing of undetected photons~\cite{Hu}. On platforms such as silicon or III-V semiconductor waveguides the integration of superconducting nanowire single photon detectors (SNSPDs)~\cite{Sprengers, Jahanmirinejad, Reithmaier, Sahin, Zhou, Kaniber, Najafi, Mattioli, Li, Pernice, Akhlaghi, Cavalier, Ferrari, Kahl, Schuck2, Schuck1, Schuck3, Beyer, Shainline, Rath, Atikian, Heeres1, Heeres2} or transition edge sensors (TESs)~\cite{Gerrits1, Calkins} has been realized. However, the integration of single-photon detectors on lithium niobate waveguides is challenging~\cite{Tanner, Hoepker}. 

TESs operate at a transition stage between a superconductive and a normal resistance state~\cite{Cabrera, Miller2}. When voltage-biased at their transition, the detector works as a micro-calorimeter, which is sensitive to temperature changes introduced by single photon absorption even at infrared wavelengths. Using their weak electron-phonon coupling, TESs made of thin-film tungsten have shown remarkable properties in terms of detection efficiency (95\,\%-98\,\%) and energy resolution (0.29\,eV-0.42\,eV)~\cite{Lita, Fukuda}. 

In this letter, we report on the first proof-of-principle evanescent single-photon detection with a transition edge sensor on a lithium niobate waveguide. This completes the toolbox for integrated quantum optics on this platform, adding integrated detection. We first show the waveguide fabrication and detector fabrication in section \ref{kap:fabrication}. 
In section \ref{kap:simulations}, we describe simulations which estimate the detection efficiency. In section \ref{kap:results} we show our experimental results including photon-number resolution up to six photons, system detection efficiency measurements for both polarizations, and first results for the energy resolution and decay time.

\section{Waveguide and detector fabrication} \label{kap:fabrication}
The waveguide fabrication starts with an 80\,nm titanium deposition on a congruent lithium niobate wafer using e-beam evaporation, followed by a positive photoresist development. Under vacuum contact lithography and subsequent wet etching, 5\,$\mu$m, 6\,$\mu$m, and 7\,$\mu$m wide titanium stripes are formed. These stripes are diffused into the lithium niobate substrate at 1060\,$^{\circ}$C creating an index gradient from 2.211 to 2.214 in TE- and 2.132 to 2.138 in TM-polarization at 1550\,nm wavelength~\cite{Edwards, Jundt, Strake}. This way both polarizations (TE and TM) can be guided with losses below 0.02\,dB/cm. A 2.5\,cm long sample consisting of 70 waveguides is cut and its end-faces are polished. To ensure the waveguide quality, loss measurements are executed using an interferometric technique described by Regener and Sohler~\cite{Regener}, where similar loss values were achieved.

\begin{figure} [h]
	\centering
	\includegraphics[width=1.0\linewidth]{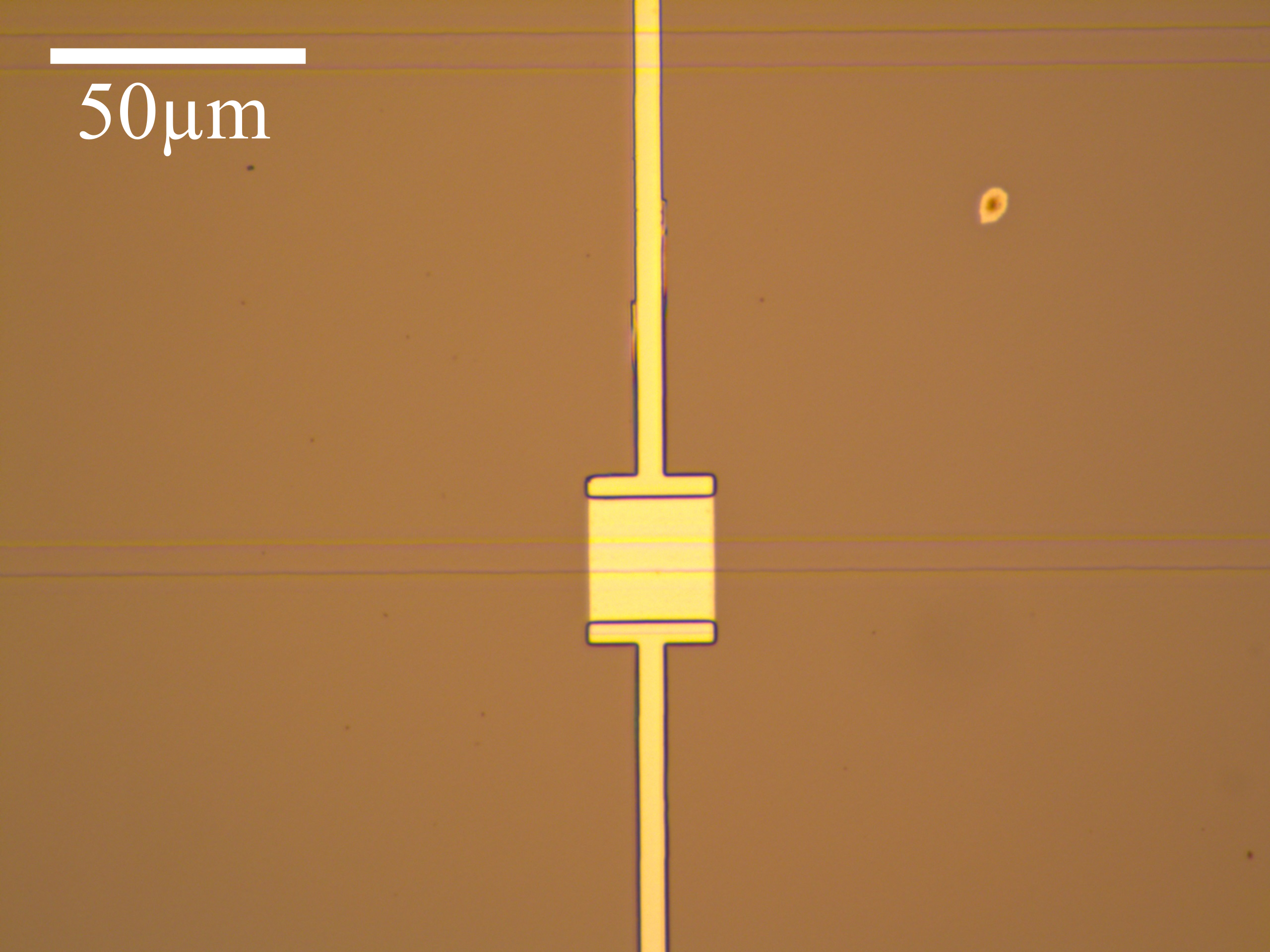} 
    \caption{Microscope image of an integrated 25\,$\mu$m x 25\,$\mu$m x 20\,nm tungsten TES on a 6\,$\mu$m wide lithium niobate waveguide.}
	\label{fig:microscopecore}
\end{figure}

After ensuring the low-loss waveguiding, the TESs are deposited. Our integrated TES devices comprise a homogeneous 20\,nm thick tungsten layer, deposited using magnetron sputtering. As 25\,$\mu$m x 25\,$\mu$m devices show high yield for fiber-coupled TESs, the same structure is chosen in this work. We place three detectors per waveguide on top of a 2\,nm amorphous silicon layer, which does not effect the optical properties of the detectors or waveguide, and additional niobium-contact pads for wire-bonding using photo lithography. A micrograph image of one device is shown in figure \ref{fig:microscopecore}.

One benefit of titanium in-diffused waveguides on lithium niobate is the mode diameter which closely matches optical fiber. With this, a single-mode fiber ferrule can be directly glued to the polished waveguide end-face using a UV-sensitive adhesive. In this device, we used standard single-mode fiber (not polarization-maintaining), in order to simplify the pigtailing process. A precise, motorized alignment setup is used to ensure good coupling between the two interfaces. A thin layer of UV-glue with a fiber-matching refractive index is deposited in between the waveguide end-face and the fiber-ferrule and symmetrically cured using a ring of UV-LEDs for pre-curing and a UV-gun to fully harden the glue. To minimize displacement while cooling the pigtailed sample inside a cryostat, the thickness of the glue layer must be minimized and a symmetric spreading of the glue is preferable. At room-temperature a theoretical maximum coupling to fiber of 92\,\% can be achieved with the given mode overlap~\cite{Montaut}.

\section{Simulations} \label{kap:simulations}
\begin{figure} [ht]
	\centering
    \subfloat[TM-polarization]{\includegraphics[width=0.49\linewidth]{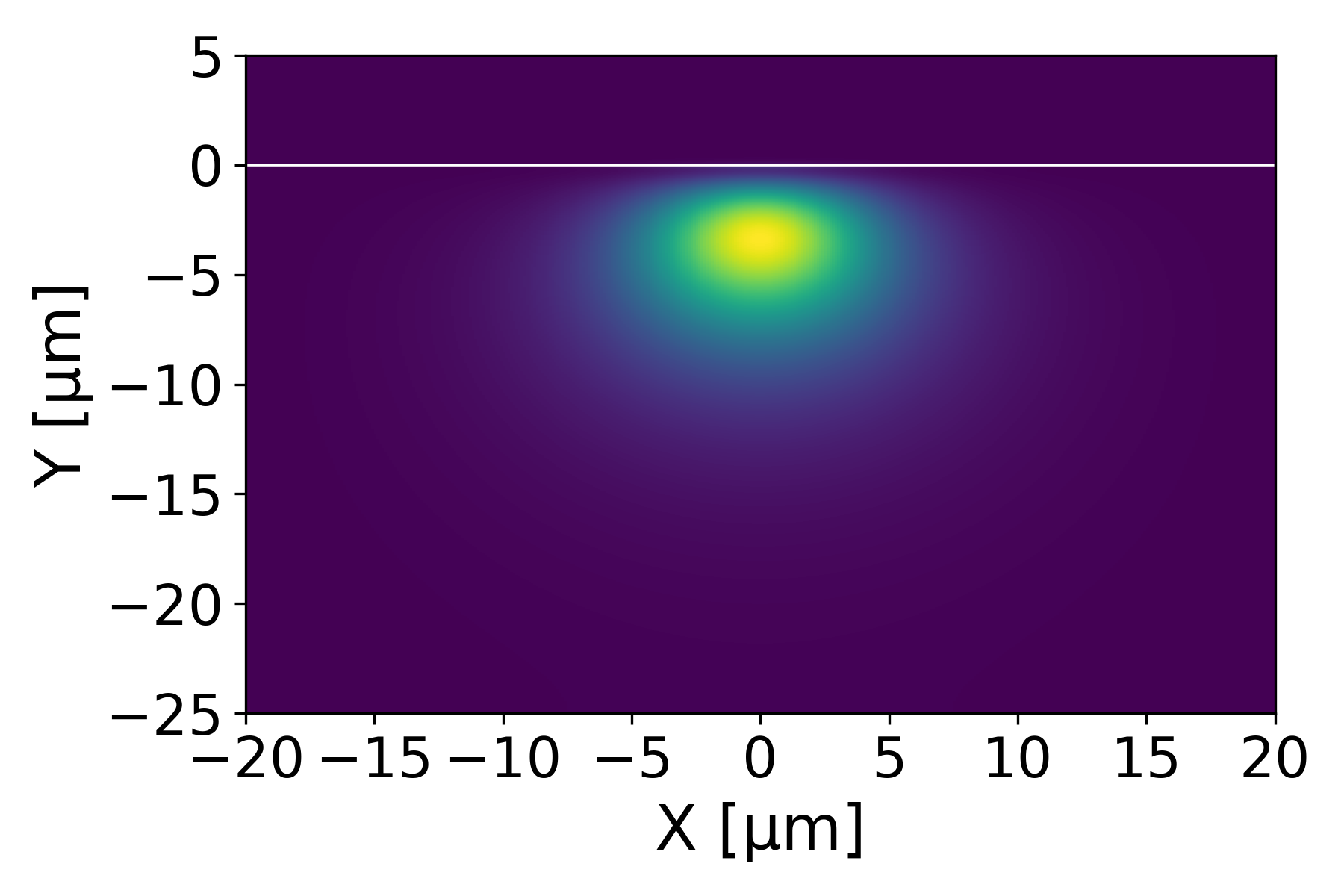}} 
    \subfloat[TE-polarization]{\includegraphics[width=0.49\linewidth]{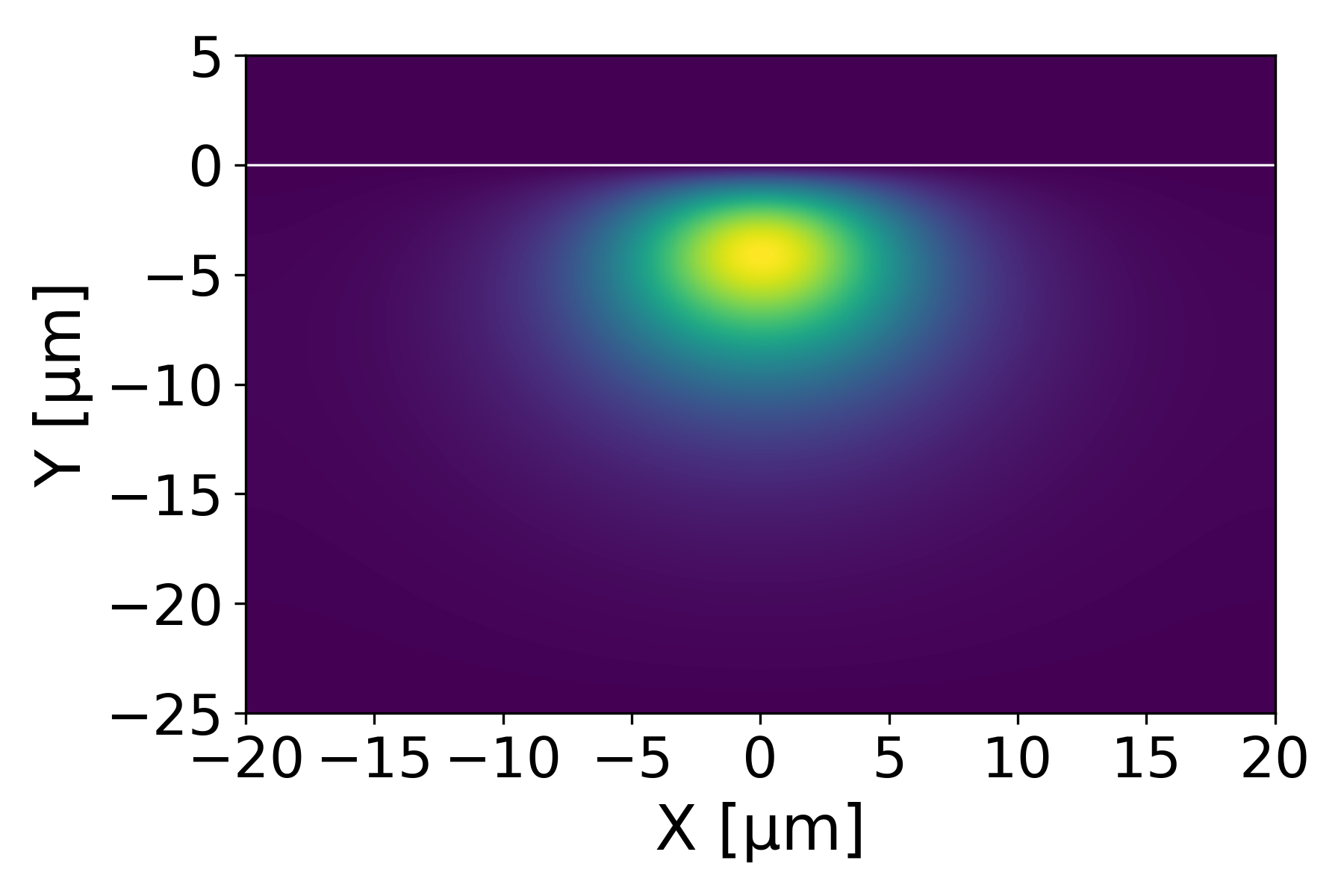}} \\
    \subfloat[TE-polarization close-up]{\includegraphics[width=0.49\linewidth]{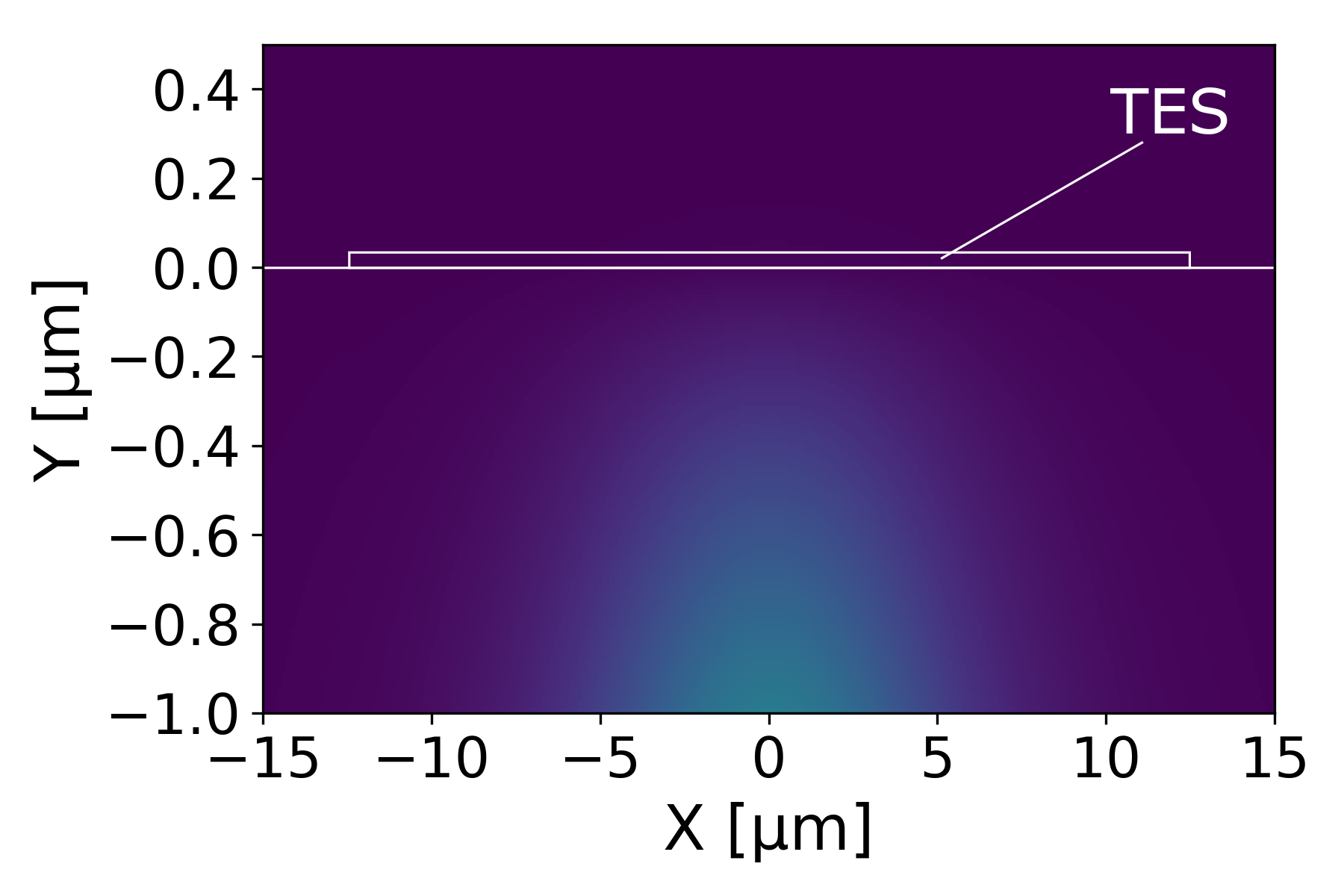}} 
    \subfloat[TE-polarization at X=0\,$\mu$m]{\includegraphics[width=0.49\linewidth]{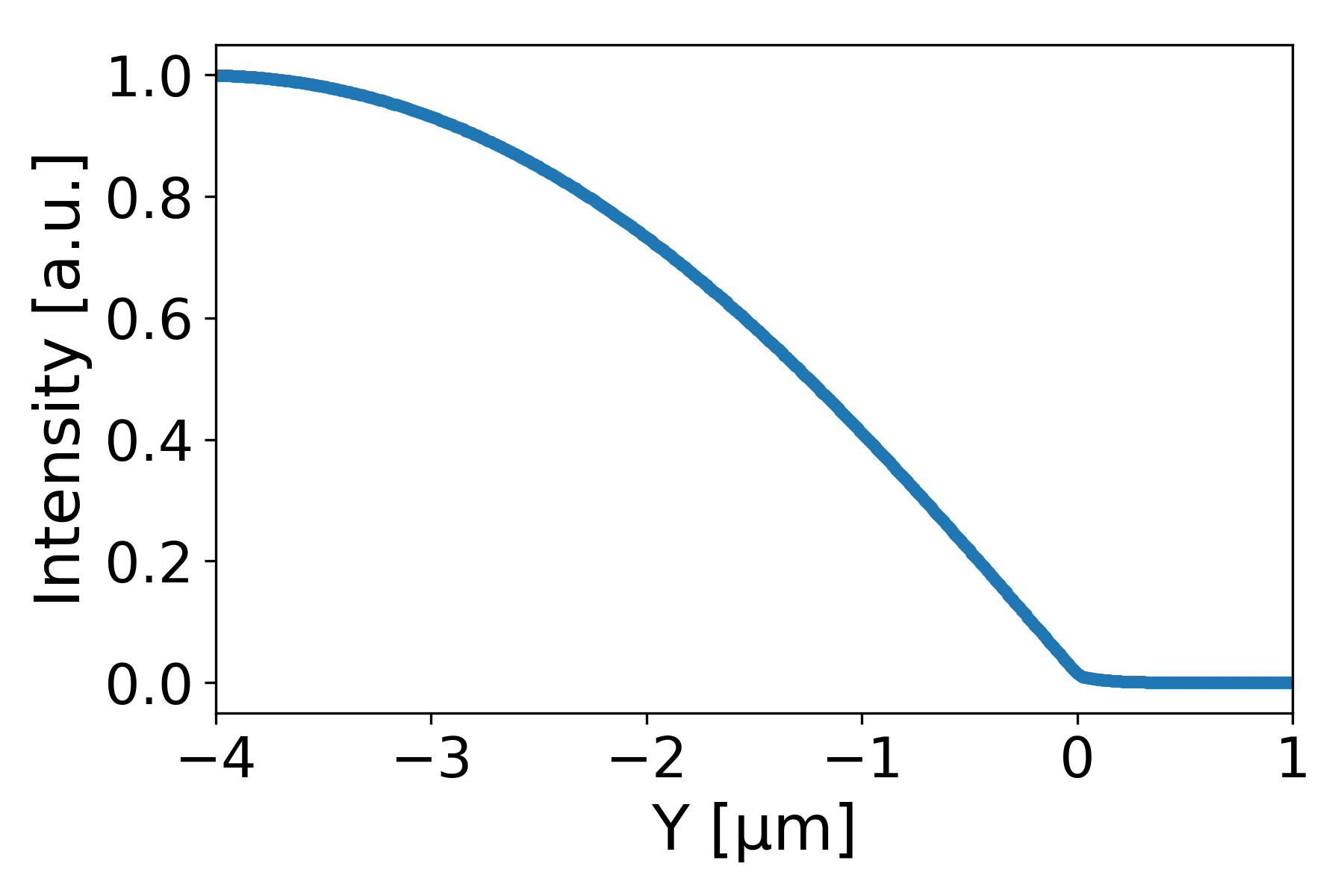}} \\
	\caption{Simulated polarization modes of a lithium niobate waveguide. (a) and (b) show the guided modes in TM- and TE-polarization. (c) shows a close-up of the surface region for the TE-mode. In (d) the mode intensity of the TE-mode is plotted in the y-direction.}
	\label{fig:modes}
\end{figure}

As the guided modes inside the waveguide were optimized to match the mode diameter of a standard single-mode fiber, the overlap of the waveguide mode and the tungsten layer is very small. Therefore, the 25\,$\mu$m x 25\,$\mu$m x 20\,nm device only sees a small part of the mode, as illustrated in figure \ref{fig:modes}. Finite-Element-Method and Finite-Difference Beam Propagation Simulations were executed using a comercial mode-solver to estimate the detector efficiency for the given structure, using the refractive index of the waveguides based on Edwards~\cite{Edwards}, Jundt~\cite{Jundt}, and Strake~\cite{Strake} as well as independently measured values for the refractive index of tungsten. From these simulations a strong polarization dependence in the absorption was calculated, with values of 1.3\,\% $\pm$0.6\,\% in TM-polarization and 0.16\,\% $\pm$0.06\,\% in TE-polarization. Although the absorption and therefore device efficiency is small, modifications to the structure can be implemented to enhance the absorption as well as multiplexing several detetors~\cite{Calkins}.

\section{Results} \label{kap:results}
\begin{figure} [h]
	\centering
	\includegraphics[width=0.97\linewidth]{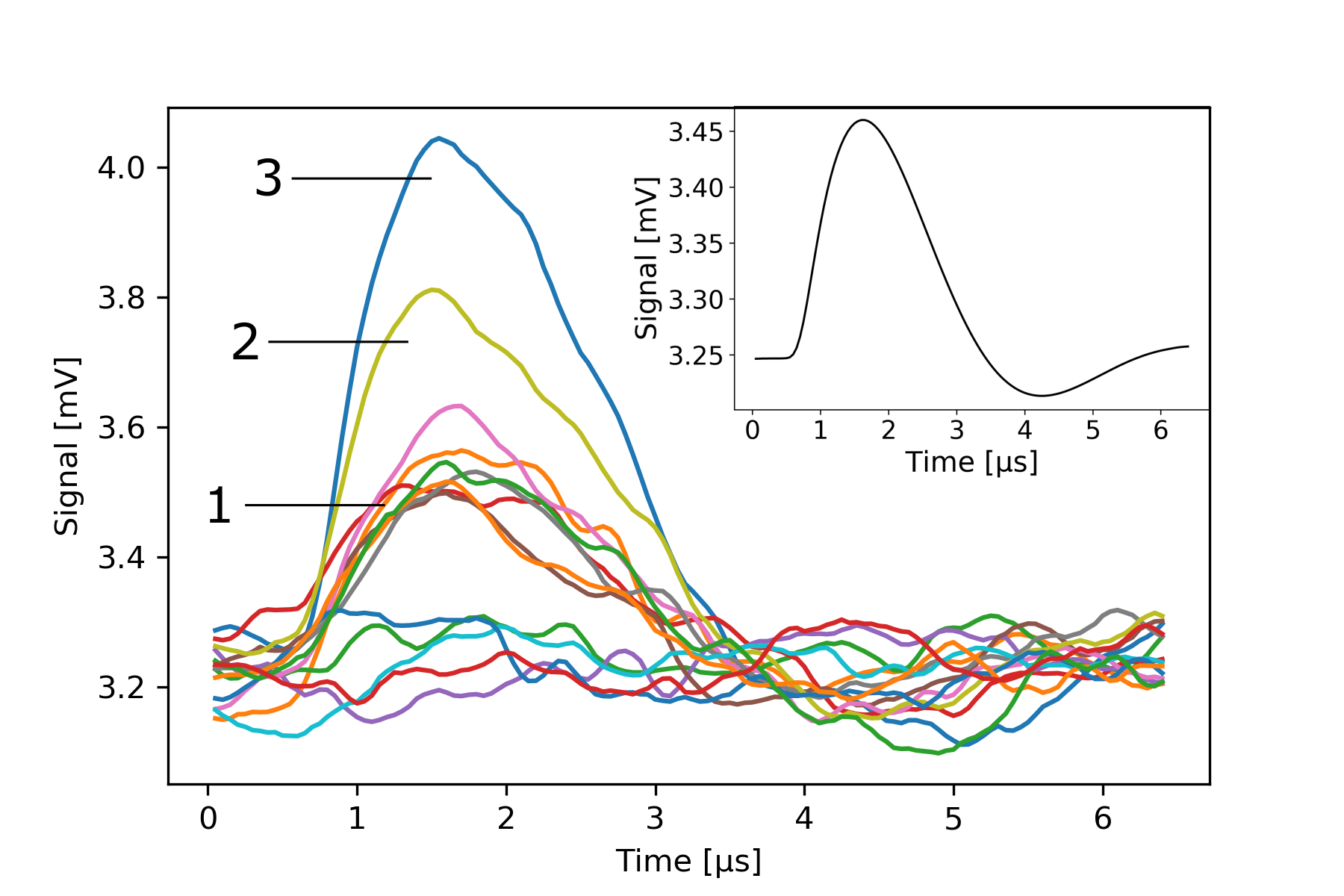}
	\caption{Measured photon traces with different peak heights corresponding to different photon numbers per pulse for an on-chip TES. The inset shows the mean photon trace.}
	\label{fig:traces}
\end{figure}

We first tested the device under flood illumination, which showed successful electrical connection and optical response for the 25\,$\mu$m x 25\,$\mu$m x 20\,nm devices~\cite{Hoepker}. Following this, the device was pigtailed and installed inside a dilution refrigerator (DR), to investigate the sensitivity to the evanescent field. The packaged device is robust and remains functional after several temperature cycles. The pigtailed fibers from the sample were spliced and connected from inside the DR to external FC/PC connectors and connected to an attenuated pulsed 1550\,nm laser. As both end-faces of the waveguide were pigtailed, stable transmission through the waveguide at cold temperatures could also be verified. An overall transmission of 43\,\% at room temperature and 8\,\% $\pm$2\,\% at 0.01\,K was measured for both polarization directions. From this, a coupling efficiency to the waveguide of 66\,\% at room temperature and 28\,\% at 0.01\,K is estimated. After optimizing the detector output, photon traces with different peak heights corresponding to different photon numbers per pulse could clearly be observed, as shown in figure \ref{fig:traces}. 

\begin{figure} [h]
	\centering
	\includegraphics[width=1.0\linewidth]{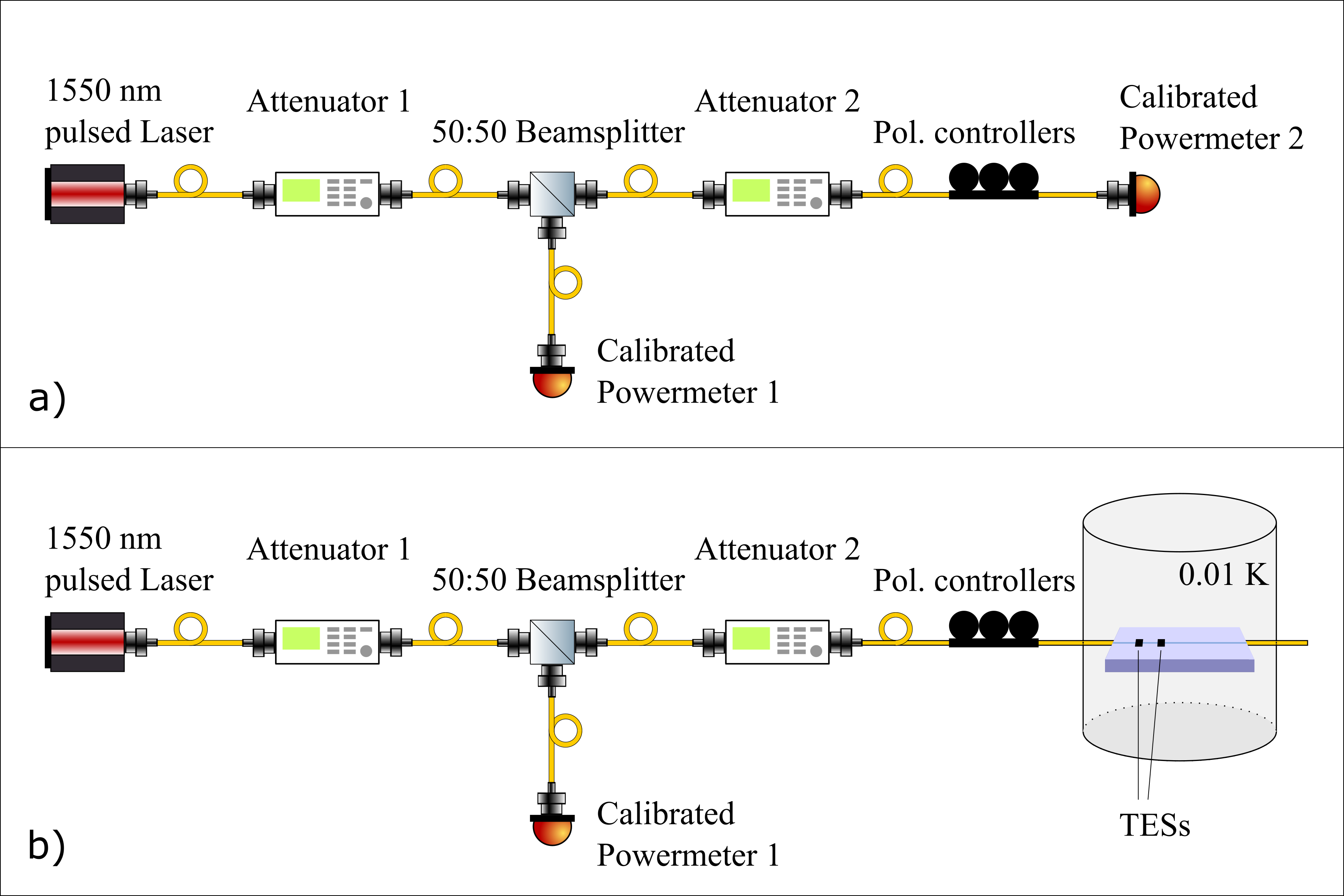}
	\caption{Experimental scheme. The splitting ratio between the two arms is measured with two calibrated powermeters as shown in the top-half. In the bottom-half the single-photon flux entering the cryostat is monitored.}
	\label{fig:detcalsetup}
\end{figure}

The detector response for two detectors on the same waveguide was measured for different polarization settings and both coupling directions. We observed a strong polarization sensitivity but a small influence of the in-coupling direction, which verifies the evanescent coupling. When measuring scattered light, one would expect a large influence on coupling direction, as the detectors are placed not symmetrically along the waveguide, as illustrated in figure \ref{fig:detcalsetup} b).

\begin{figure} [ht!]
	\centering
    \subfloat[TM-polarization]{\includegraphics[width=0.97\linewidth]{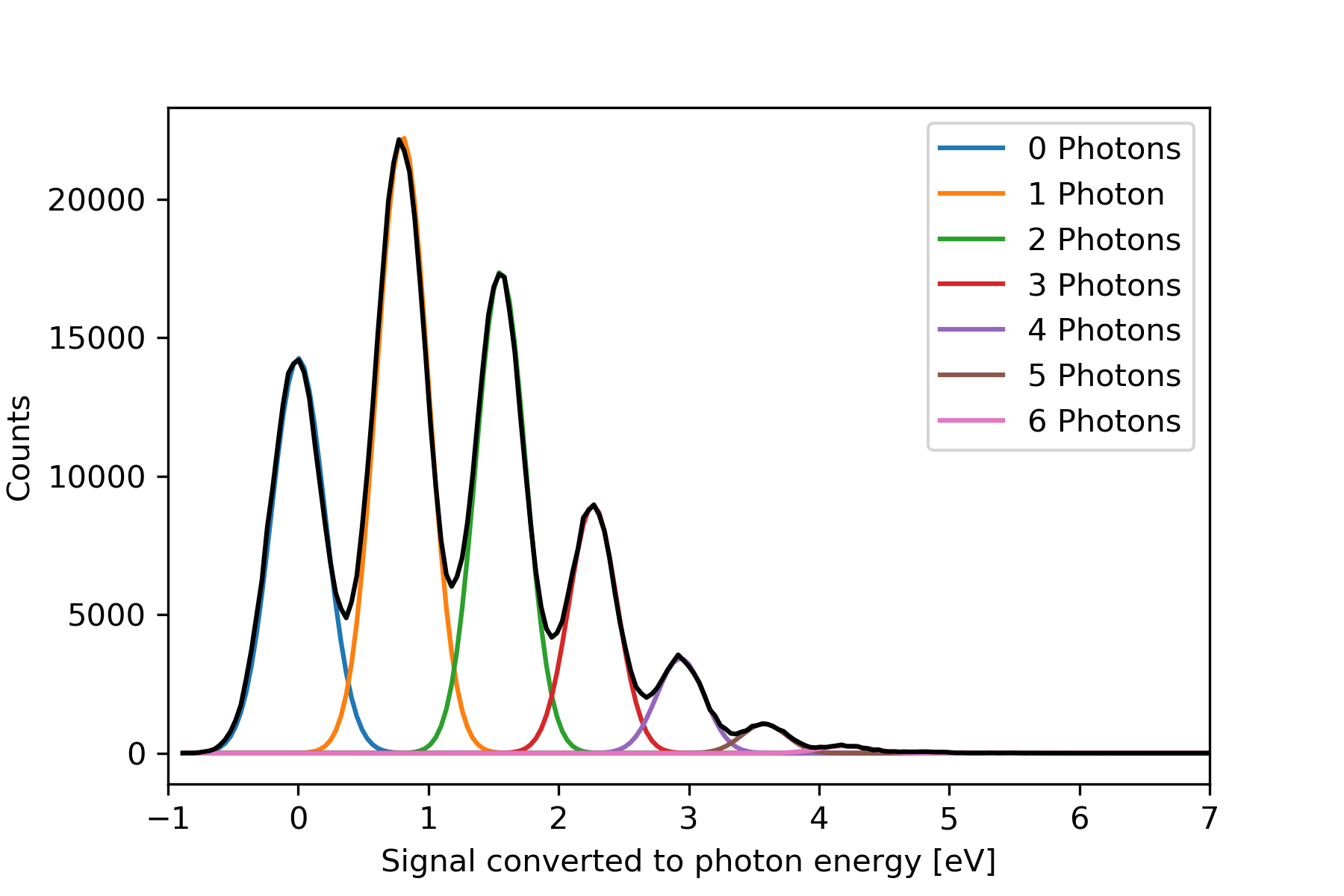}} \\
    \subfloat[TE-polarization]{\includegraphics[width=0.97\linewidth]{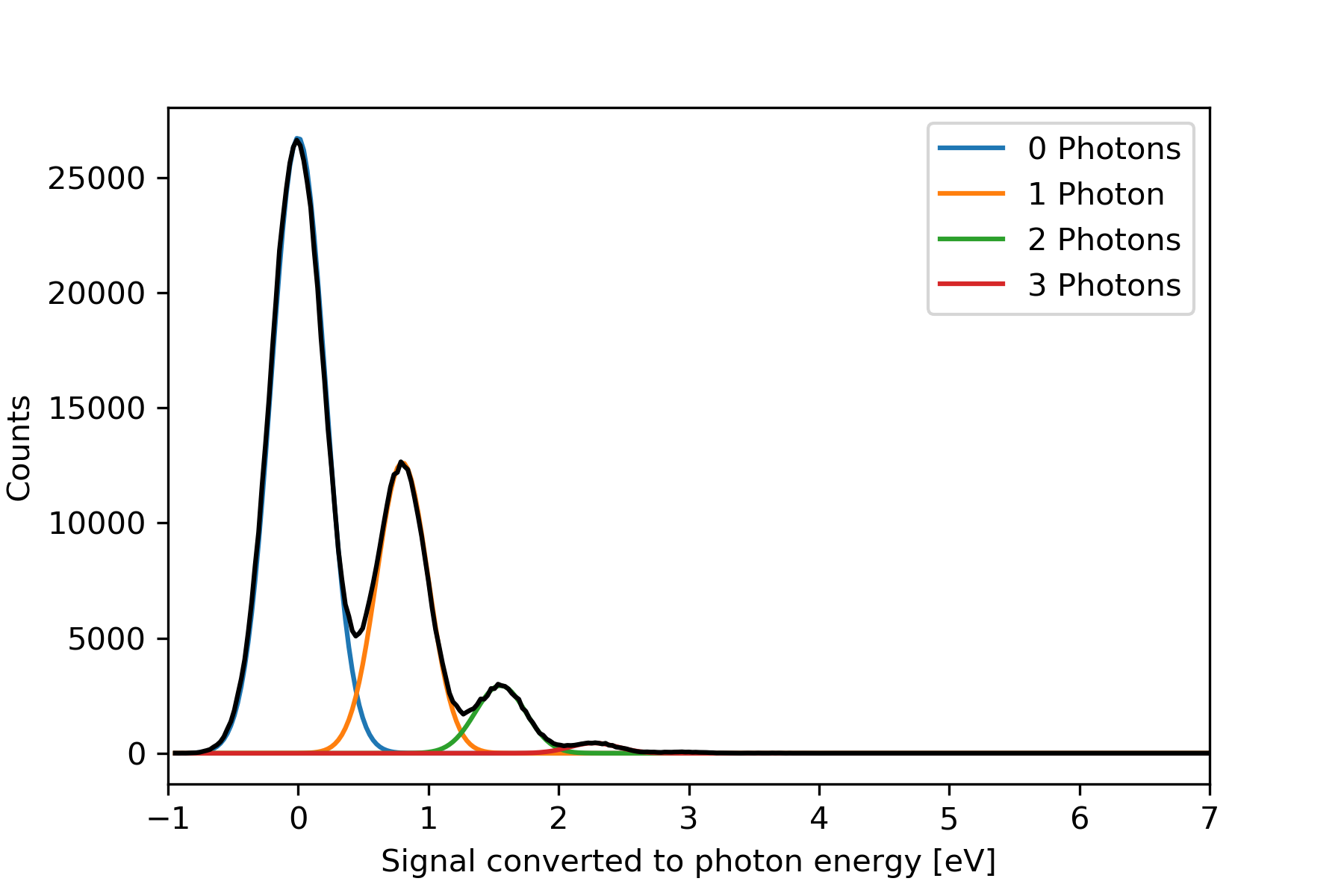}}
	\caption{Histograms of the evaluated photon traces converted to the underlying photon energies for one detector and gaussian fitting for the individual peaks.}
	\label{fig:gauss}
\end{figure}

From the measured photon traces of each detector and for each polarization setting, a mean photon trace can be calculated as a template \cite{Fixsen}. From the convolution of each trace and the template, we construct histograms for individual detectors, polarization setting, and in-coupling direction, as shown in figure \ref{fig:gauss}. Also, an average (1/e) decay time of 1.6\,$\mu$s can be calculated from a mean photon trace. By using a Gaussian fit or by manually setting individual thresholds, the histogram can be used to calculate a mean photon number. In addition, we can determine the energy resolution of the detectors from the histogram using the mean photon energy of the zero and one photon peak and its FWHM. We found an average energy resolution of 0.33\,eV $\pm$0.05\,eV, which is similar to other platforms.

We measured the expected mean photon number using a splitting-ratio method as illustrated in figure \ref{fig:detcalsetup}. In a first step, the splitting ratio between the two arms is investigated, using low attenuation on Attenuator 1 and high attenuation on Attenuator 2, which leads to high transmission at the calibrated Powermeter 1 and a lower but still measurable signal at the calibrated Powermeter 2. Next, Powermeter 2 is exchanged with a fiber that feeds into the cryostat and is pigtailed at the waveguide end-face. Using a high attenuation on Attenuator 1, while keeping Attenuator 2 fixed, reduces the output in both arms. Measuring the output at Powermeter 1 can be used to calculate the single-photon flux in the other arm.

From this, as well as an expected mean photon number, the maximum and minimum system detection efficiency (SDE) is calculated, depending on the polarization. This gives 0.23\,\% $\pm$0.04\,\% and 0.06\,\% $\pm$0.01\,\% for one detector and 0.19\,\% $\pm$0.04\,\% and 0.07\,\% $\pm$0.01\,\% for a second tested detector. The in-coupling direction has no influence on the SDE within the measurement accuracy. 

The TES response was optimized by changing the polarization setting and monitoring the mean photon number using an $in$ $situ$ histogramming. From our room-temperature loss measurements and simulations the maximum SDE was related to TM-polarization and the minimum SDE to TE-polarization. Further experiments using polarization maintaining fibers should be executed to verify this assumption. From transmission measurements through the pigtailed waveguide, we subtract coupling losses into the chip and calculate a corrected maximum detection efficiency of 0.7\,\%  per detector. Compared to simulations, the lower efficiency can be attributed to a simplified model used to calculate the absorption in the simulations and to the polarization setting accuracy.

\section{Conclusion} \label{kap:conclusion}
In this paper, we demonstrated the first proof-of-principle detection of single photons using TESs on lithium niobate waveguides. From the measured photon traces a histogram was calculated to confirm the photon-number resolution of the detectors. Light was coupled in from both directions into two detectors on the same waveguide and a strong polarization sensitivity was observed. Also, values for the average response decay time and energy resolution were calculated. 

As an outlook, the detection efficiency can be significantly increased by changing the detector and/or waveguide geometry. As already shown on silica-on-silicon waveguides, additional tungsten fins (metallic absorptive strips deposited on the waveguide to conduct heat to the TES) can be used to increase the interaction length with the waveguide~\cite{Calkins}. This can increase the evanescent coupling by at least one order of magnitude. Furthermore, the system detection efficiency can be enhanced by detector multiplexing.

\section*{Acknowledgements}
Funded by the Deutsche Forschungsgemeinschaft (DFG, German Research Foundation)
- Projektnummer 231447078 - TRR 142. This is a contribution of NIST, an agency of the U.S. government, not subject to copyright. 

\bibliography{Bibo101018}

\end{document}